\documentclass{jltp}

\usepackage{graphicx}

\title{Superfluid to Mott insulator transition \\ in one, two, and three dimensions}

\author{M. K{\"o}hl, H. Moritz, T. St{\"o}ferle, C. Schori, T. Esslinger}

\address{Institute of Quantum Electronics\\ ETH Z\"{u}rich
H\"{o}nggerberg\\ CH--8093 Z\"{u}rich, Switzerland}

\runninghead{M. K{\"o}hl {\it et al.}}{Superfluid to Mott
insulator transition in 1D, 2D, and 3D}

\begin{document}

\maketitle

\begin{abstract}
We have created one-, two-, and three-dimensional quantum gases
and study the superfluid to Mott insulator transition.
Measurements of the transition using Bragg spectroscopy show that
the excitation spectra of the low-dimensional superfluids differ
significantly from the three-dimensional case.


\end{abstract}

\section{LOW-DIMENSIONAL QUANTUM SYSTEMS}
A low dimensional gas can be created in a trap when the confining
potential restricts the motion of the particles to one or two
dimensions with the other motional degrees of freedom being frozen
out. Quantum mechanically this implies that the particles occupy
only the ground state in the restricted directions, i.e. both, the
thermal energy $k_B T$ and the interaction energy $\mu$ have to be
much smaller than the energy level spacing. In general,
low-dimensional systems exhibit increased quantum fluctuations of
the phase. For a {\it homogeneous} 2D gas these phase fluctuations
allow Bose-Einstein condensation only at zero temperature and for
a homogeneous 1D gas no Bose condensation is possible at all. For
{\it trapped} low-dimensional gases the Bose-Einstein phase
transition takes place at finite temperatures for both,
1D\cite{Ketterle1996} and 2D\cite{Petrov2000b}, but the
fluctuating phase alters the properties of the gas.
Low-dimensional quantum systems exhibit a wealth of fascinating
phenomena whose explanations go beyond the mean-field
description\cite{Giamarchi2004}. For example, in one dimension
bosons acquire fermionic properties at low densities and strong
interactions, forming the so-called Tonks-Girardeau
gas\cite{Girardeau1960}. For two dimensions a
Berezinskii-Kosterlitz-Thouless transition is predicted resulting
in the spontaneous formation of vortices from thermal
excitations\cite{Berezinskii1971}.

A one-dimensional gas can be realized in a cigar shaped harmonic
trapping geometry characterized by the frequencies
$\omega^{1D}_\perp$ in the two strongly confining axes and
$\omega^{1D}_z$ in the weakly confining axis. A two-dimensional
trap can be realized in a disk shaped geometry with one strongly
confining axis characterized by $\omega^{2D}_z$ and two weakly
confined directions by $\omega^{2D}_\perp$. The conditions for
achieving the low-dimensional quantum regime then read
\begin{eqnarray}
k_B T, \mu \ll  \Big \{ \begin{array}{ll}
  \hbar \omega^{1D}_\perp & \textrm{for 1D;} \\
  \hbar \omega^{2D}_z & \textrm{for 2D.}
\end{array}
\end{eqnarray}
A one-dimensional trapped Bose gas in the weakly interacting
regime was recently created and studied\cite{Moritz2003}. This
experiment revealed the distinctively different excitation
spectrum of a one-dimensional quantum system as compared to three
dimensions. Previous studies accessed a regime, where a Bose
condensate with $\mu \leq \hbar \omega_\perp^{1D}$ coexisted with
a three-dimensional thermal cloud\cite{Gorlitz2001,Schreck2001}.
Also Bose gases were created in optical lattices with significant
tunnelling between the one-dimensional
subsystems\cite{Greiner2001b}. Theoretical understanding of the
homogeneous one-dimensional Bose gas was pioneered by Lieb and
Liniger in the 1960s\cite{Lieb1963} and recently adapted to atom
traps\cite{Olshanii1998,Ho1999,Petrov2000a,Kagan2000,Girardeau2001,Menotti2002}.
A fundamental feature of the one-dimensional Bose gas is that its
ground state energy and excitation spectrum are determined by a
single parameter $\gamma$ over the whole range from weak to strong
interactions. $\gamma=\frac{m g_{1D}} {\hbar^2 n_{1D}}$ is the
ratio between interaction energy and kinetic energy ($m$: atomic
mass, $g_{1D}$: 1D coupling constant, $n_{1D}$: 1D density). The
transition from  weak to  strong interactions can be studied by
applying a lattice potential along the symmetry axis of the 1D
gas\cite{Stoferle2004}.

Quasi-homogeneous two-dimensional quantum gases have been studied
for several years using thin films of liquid Helium on a
surface\cite{Safonov1998}. With ultracold alkali atoms quantum
degenerate atomic gases in two dimensions have been realized in a
standing wave laser field\cite{Kasevich2001}. Moreover, the regime
with $\mu\leq\hbar \omega^{2D}_z$ has been studied using $^{23}$Na
in a disk shaped optical trap where features of two-dimensional
condensate expansion were observed\cite{Gorlitz2001}. With
$^{133}$Cs a quantum degenerate sample has been prepared in an
optical trap close to a surface\cite{Rychtarik2003} and in
extremely fast rotating condensates the two-dimensional regime has
also been approached\cite{Schweikhard2004}. In this paper we
present experiments with Bose gases fully in the 2D regime with
$k_BT/\hbar \omega^{2D}_z<6 \times 10^{-3}$ and $\mu/\hbar
\omega^{2D}_z<0.1$. The gas is subject to a periodic potential
along its weakly confined directions, which allows us to tune the
gas from a weakly interacting superfluid to a strongly correlated
Mott insulator.

\section{SUPERFLUID TO MOTT INSULATOR TRANSITION}
Bosons which are localized in the minima of a periodic potential
may be described by the Bose-Hubbard
Hamiltonian\cite{Fisher1989,Jaksch1998}:
\begin{eqnarray}
H=-\widetilde{J} \sum_{i,j} \hat{a}^\dagger_{j} \hat{a}_i+ \sum_i
\epsilon_i n_i+\frac{U}{2} \sum_i n_i (n_i-1).
\end{eqnarray}
$\widetilde{J}$ denotes the hopping matrix element between
neighboring lattice sites that determines the rate of which a
particle disappears from lattice site $i$ and tunnels to the
adjacent lattice site $j$ ($\hat{a}_i$ and $\hat{a}^\dagger_i$ are
the annihilation and creation operators for an atom at lattice
site $i$, respectively). The total tunnelling rate $J$ includes
the possibly anisotropic tunnelling to all next neighboring
lattice sites:
$J=2(\widetilde{J}_x+\widetilde{J}_y+\widetilde{J}_z)$.
$\epsilon_i$ characterizes the inhomogeneity of the atom trap.
$n_i$ is the occupation number of lattice site $i$ and $U$ is the
onsite-interaction energy between two bosons at the same lattice
site. The Bose-Hubbard Hamiltonian exhibits a quantum phase
transition between the superfluid state and a Mott insulator for a
critical ratio $(U/J)_c$ which has been observed in an optical
lattice, where the periodic potential is created by standing wave
laser fields\cite{Greiner2002a}. The intensity of the laser fields
determines the value of $U/J$. According to a mean-field
calculation one expects $(U/J)_c=5.8$\cite{Fisher1989,Jaksch1998}.
This approximation is valid for two and three dimensions, but for
one spatial dimension stronger quantum fluctuations of the phase
must be taken into account\cite{Kuehner1998}. Quantum Monte-Carlo
simulations show\cite{Batrouni2002} that these lower the onset of
the Mott insulating phase to $(U/J)_c \simeq 1.8$. We have
recently obtained experimental evidence for these
fluctuations\cite{Stoferle2004}.
\begin{figure}[htbp]
\begin{center}
  \includegraphics[width=.8\columnwidth,clip=true]{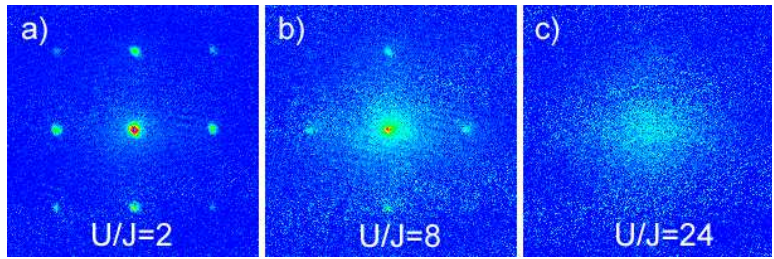}
  \caption{Quantum phase transition from a superfluid to a Mott insulator in three dimensions.
  The absorption images are taken after 25 ms of ballistic expansion. The interference peaks
  in a) are separated by a  momentum of $2\hbar k$, where $k=2 \pi/\lambda$ is the wave vector
  of the laser creating the optical lattice.}
  \label{fig1}
\end{center}
\end{figure}

One signature of the phase transition can be observed by studying
ballistic expansion of the atoms from the lattice. For strong
tunnelling, the system is superfluid, and the wave function of an
atom is delocalized over many lattice sites. The absorption image
shows interference peaks with a periodicity according to the
lattice vector (see Figure~\ref{fig1}a). In contrast, for strong
interactions in the Mott insulating phase the atoms are localized
within a single potential well and have no definite phase relation
to the next potential well. Therefore, no interference pattern
develops during the expansion (see Figure~\ref{fig1}c).

\section{EXPERIMENTAL SETUP}
\subsection{Bose condensates in an optical lattice}
In the experiment, we collect up to $2 \times 10^9$ $^{87}$Rb
atoms in a vapor cell magneto-optical trap. After polarization
gradient cooling and optical pumping into the $|F=2, m_F=2\rangle$
hyperfine ground state the atoms are captured in a magnetic
quadrupole trap. After magnetic transport of the trapped atoms
over a distance of 40\,cm the magnetic trapping potential is
converted into the harmonic and elongated potential of a QUIC
trap\cite{Esslinger1998}. Subsequently, we perform radio frequency
induced evaporation of the cloud over a period of 25\,s. After
evaporation we observe almost pure Bose-Einstein condensates of up
to $1.5\times 10^5$ atoms. Following the condensation we
adiabatically change the trapping geometry to an approximately
spherical symmetry with trapping frequencies of $\omega_x=2 \pi
\times 18$\,Hz, $\omega_y=2 \pi \times 20$\,Hz, and $\omega_z=2
\pi \times 22$\,Hz. This reduces the peak density by a factor 4
and allows us to load the optical lattice more uniformly.

\begin{figure}[htbp]
\begin{center}
  \includegraphics[width=.8\columnwidth,clip=true]{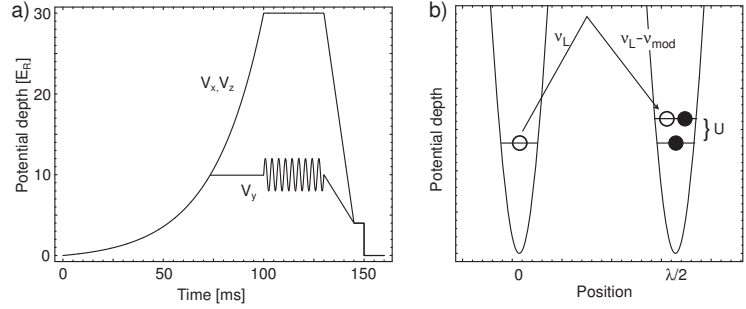}
  \caption{a) Experimental sequence to perform Bragg spectroscopy on atoms in an optical lattice.
  The figure illustrates an experiment with a one-dimensional gas. For a two- or three-dimensional
  gas the curves are accordingly different (see text). b) Scheme for resonant excitation of a Mott
  insulator by Bragg spectroscopy (not to scale).}
  \label{fig2}
\end{center}
\end{figure}

The optical lattice is formed by three retro-reflected laser
beams. Each beam is derived from a laser diode at a wavelength of
$\lambda=826$\,nm. At the position of the condensate the beams are
circularly focused to $1/e^2$-radii of $120$\,$\mu\textrm{m}$ ($x$
and $y$ axes) and $105$\,$\mu\textrm{m}$ ($z$). The three beams
possess mutually orthogonal polarizations and their frequencies
are offset with respect to each other by several ten MHz. We
stabilize the lasers to a high-finesse Fabry-Perot cavity, thereby
reducing their line width to $\sim 10\,\textrm{kHz}$. In order to
load the condensate into the ground state of the optical lattice,
the intensities of the lasers are slowly increased to their final
values using an exponential ramp with a time constant of
$25\,\textrm{ms}$ and a duration of $100\,\textrm{ms}$ (see
Fig.~\ref{fig2}). The resulting optical potential depths
$V_{x,y,z}$ are proportional to the laser intensities and are
conveniently expressed in terms of the recoil energy
$E_R=\frac{\hbar^2 k^2}{2 m}$ with $k=\frac{2 \pi}{\lambda}$ and
the atomic mass $m$.

\subsection{Preparation of low-dimensional quantum gases}
Using the optical lattice we realize one-, two-, and
three-dimensional quantum gases. To prepare an array of
one-dimensional tubes, two lattice axes are ramped to a fixed
value $V_\perp \equiv V_x=V_z=30\,E_R$. In this configuration the
transverse tunnelling rates $J_x$ and $J_z$ are small and
contribute only a correction of the order of $J_{x,z}/\mu\ll 1$ to
the 1D characteristics of the individual tubes, where $\mu$ is the
chemical potential of the sample. To study the superfluid to Mott
insulator transition in this system we apply an additional optical
lattice along the symmetry axis with a potential depth $V_y \ll
V_\perp$. Upon variation of $V_y$ the state of the system changes
from superfluid to Mott insulating. Experimentally we observe the
same behavior of the system for the following preparation schemes:
(i) when ramping all lattice laser beams at the same time, (ii)
when first preparing a one-dimensional gas and subsequently
ramping up $V_y$ and (iii) when creating a three-dimensional
lattice first and subsequently lower $V_y$ to the desired value.

We prepare two-dimensional quantum gases by applying one strong
lattice laser ($V_z=30\,E_R$) to freeze the atomic motion in the
z-direction only. The two other laser beams are kept at lower
levels with $V_x=V_y$. For the three-dimensional situation we use
the same power in all three laser beams ($V_x=V_y=V_z$).

\section{BRAGG SPECTROSCOPY IN AN OPTICAL LATTICE}
\subsection{Principle of the measurement}
The excitation spectrum reveals fundamental properties of the
quantum state of the system. We study the excitation spectrum by
two-photon Bragg spectroscopy\cite{Stoferle2004,Stenger1999}
employing amplitude modulation of the lattice potential. For the
one-dimensional situation we modulate the weak axial potential,
for the two-dimensional situation we modulate one of the two weak
standing wave laser fields. The modulated lattice potential takes
the form $V_{y}(y,t) = V_{y,0}(1 + A_{mod}\sin{(2\pi\nu_{mod}t)})
\sin^2(ky)$. The modulation with amplitude $A_{mod}$ and frequency
$\nu_{mod}$ generates two sidebands with frequencies $\pm
\nu_{mod}$ relative to the lattice laser frequency $\nu_L$ which
define the energy $h\nu_{mod}$ of the excitation. Due to the Bragg
condition atoms scattering two photons receive a momentum transfer
of $0\hbar k$ or $\pm 2\hbar k$.

In contrast to applying a potential gradient across the
lattice\cite{Greiner2002a}, this method is not susceptible to
effects like Bloch oscillations, dynamical instabilities, and
Zener tunnelling which occur for low axial lattice depths.
Furthermore, the excitation energy is precisely determined.

After the excitation, the experimental sequence continues by
ramping down the lattice potentials linearly in 15\,ms to
$V_{x,y,z}=4\,E_R$ where the atoms are able to tunnel again in all
three dimensions between the sites of the lattice. To allow for
re-thermalization of the system, the atoms are kept at this
lattice depth for $5\,\textrm{ms}$. Then all optical and magnetic
potentials are suddenly switched off. The switch-off time for the
optical lattice is 5\,$\mu$s and for the magnetic trap
300\,$\mu$s. The resulting matter wave interference pattern is
detected by absorption imaging after $25\,\textrm{ms}$ of
ballistic expansion. The width of the central momentum peak is
taken as a measure of how much energy has been deposited in the
sample by the excitation. If the energy increase is small, the
peak is well fitted by a bimodal distribution. For resonant
excitation there is only a single gaussian component, reflecting
that the temperature of the atoms has significantly increased. To
be independent of the shape of the peak we use the full width at
half maximum (FWHM) as a measure of the introduced energy.
Although this underestimates small energy increases, the important
resonances and features of the spectra are well shown.

\begin{figure}[htbp]
\begin{center}
  \includegraphics[width=.9\columnwidth,clip=true]{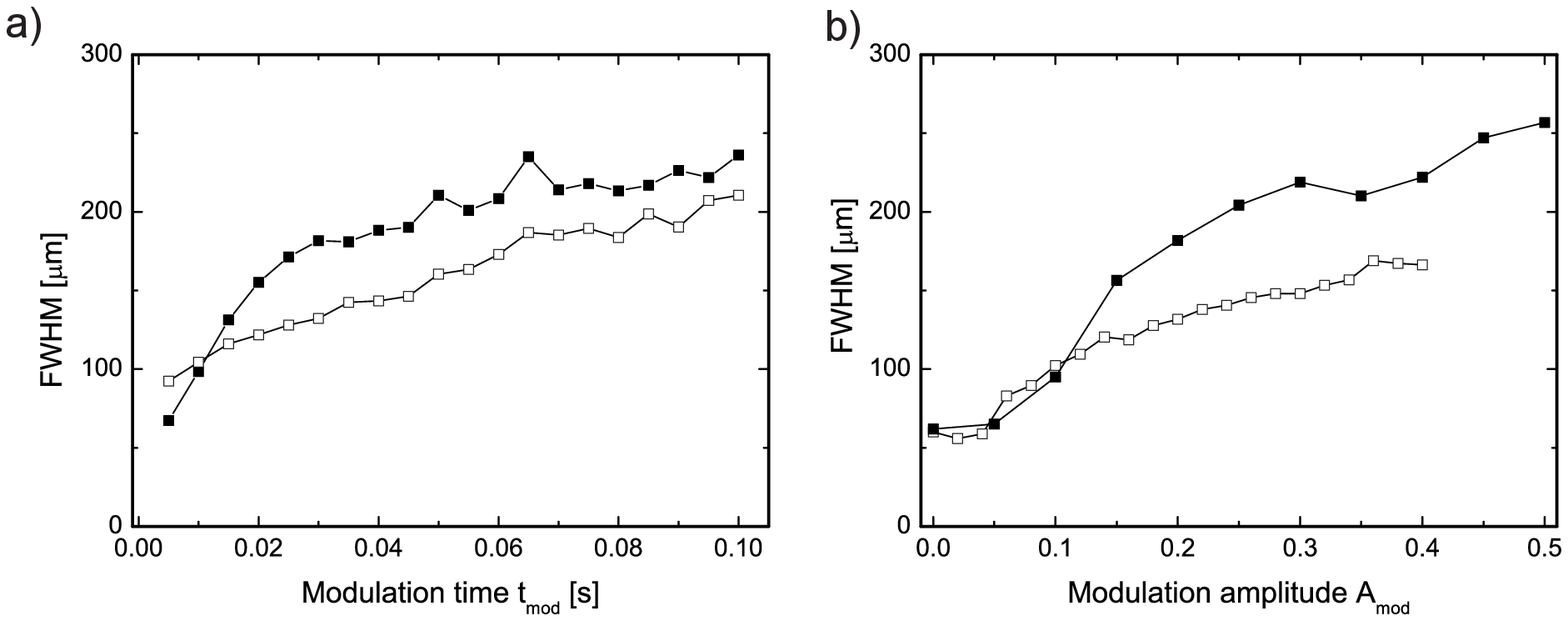}
  \caption{Linearity of the excitation scheme for a
  one-dimensional gas. Filled symbols correspond to the superfluid
  regime, with $V_y=4\,E_R$, open symbols to the
  Mott insulating regime with $V_y=10\,E_R$. All data where
  taken close to the peak of the energy absorption at a modulation frequency of
  $\nu_{mod}=1890$\,Hz.
  a) Modulation with constant modulation index $A_{mod}=0.2$.
  b) Modulation with fixed duration of $0.03$\,s.}
  \label{fig4}
\end{center}
\end{figure}

We have chosen a duration $t_{mod}=30\,\textrm{ms}$ and amplitude
$A_{mod}=0.2$ of the modulation such that the resulting excitation
of the condensate does not exhibit saturation effects for all
measurements presented here (see Fig. \ref{fig4}). We have
experimentally verified that all atoms remain in the lowest Bloch
band by adiabatically switching off the lattice
potentials\cite{Greiner2001b} after the modulation. For a thermal
cloud of atoms loaded into an optical lattice we do not observe an
energy increase by the modulation.

\subsection{Results}
\begin{figure}[htbp]
\begin{center}
  \includegraphics[width=.45\columnwidth,clip=true]{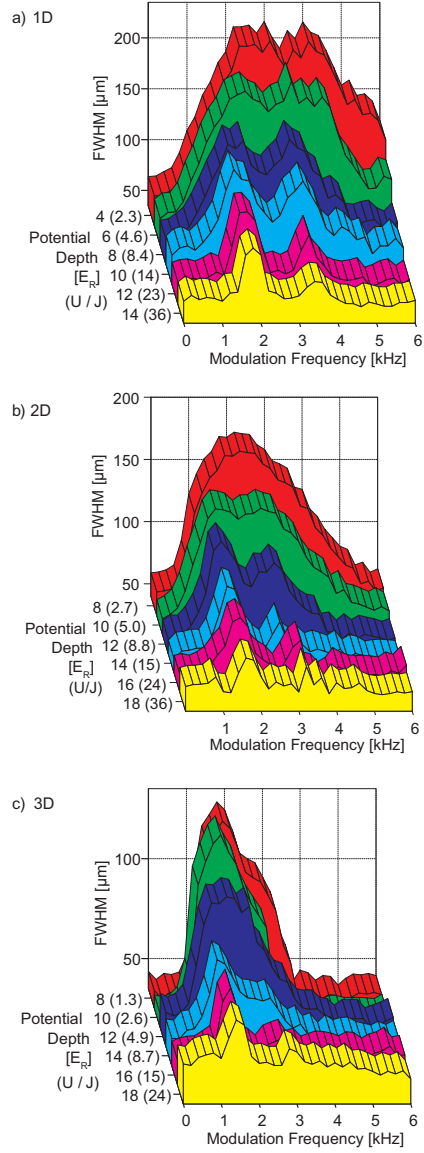}
  \caption{The measured excitation spectra of the superfluid to Mott insulator transition.
  a) An array of 1D gases ($V_x=V_z=30\,E_R$).
  b) Superfluid to Mott insulator transition in 2D
  ($V_z=30\,E_R$). c) Superfluid to Mott insulator transition in 3D
  ($V_x=V_y=V_z$). The ratios $U/J$ given in brackets are calculated
  numerically. The data for a) and c) are taken from reference\cite{Stoferle2004}.}
  \label{fig3}
\end{center}
\end{figure}

The fundamental change in the excitation spectrum when the system
undergoes the quantum phase transition from a superfluid to a Mott
insulator can be seen in Figure~\ref{fig3}: The broad continuum of
the superfluid ($U/J\ll 5.8$) contrasts with the discrete spectrum
of the Mott insulator ($U/J\gg 5.8$). Figure~\ref{fig3}a displays
the 1D situation with $V_\perp=30\,E_R$ and includes a series of
spectra for different values of $U/J$. The transition in a
two-dimensional gas ($V_z=30\,E_R$) is shown in b). Finally,
Figure~\ref{fig3}c represents the three-dimensional case with
$V_x=V_y=V_z$.
One surprising feature is that we can excite the superfluid with
our scheme at large $h\nu_{mod}$ contrary to predictions for the
weakly interacting superfluid in an optical lattice formed by a
single standing wave\cite{Menotti2003}. In our experiment strong
interactions lead to a significant quantum depletion. It amounts
to $\approx 50\%$ for the 1D configuration with
$U/J=2.3$\cite{Kraemer2003}. This has been studied quantitatively
in a measurement of the coherence properties of the
system\cite{Stoferle2004}. Therefore this parameter may not be
regarded small as in standard Bogoliubov theory and higher order
excitations should be taken into account\cite{Hugenholtz1959}. In
combination with the broken translational invariance in the
inhomogeneous trap, this could explain the non-vanishing
excitation probability observed in the experiment at high
energies\cite{Buechler2003}.

In the superfluid regime we obtain spectra which differ
significantly from previous results for three
dimensions\cite{Greiner2002a}, since the superfluid excitations
decrease at higher energies. This decrease is rather slow for the
1D gas but becomes more pronounced for the 2D and 3D gases. Our
excitation scheme does not induce dephasing that occurs when the
strongly interacting condensate is accelerated near the edge of
the Brillouin zone\cite{Bronski2001}. This might cause the
broadening and the background in the tilted lattice
experiments\cite{Greiner2002a} at high energies. The width of the
superfluid spectra for the 1D gas is on the same order as twice
the width of the lowest band for Bogoliubov
excitations\cite{Kraemer2003b}.

We observe the appearance of the discrete structure, which is
characteristic for the Mott insulating phase, between $U/J= 4$ and
$U/J= 8$. Above $U/J= 20$ there is no more background due to the
vanishing superfluid component. However, the finite size and the
inhomogeneity of the trap prohibit a sharp transition, so that the
fraction of Mott insulating atoms increases gradually with $U/J$.
In the Mott insulating phase we find the first resonant peak for
all data sets close to the calculated value of $U$. A second peak
appears at $(1.91\pm 0.04)$ times the energy of the first
resonance. We attribute this resonance to defects where lattice
sites with $n=1$ atom next to sites with $n=2$ atoms are being
excited. For the 1D and 2D system (Fig.~\ref{fig3}a and b) a much
weaker resonance appears at $(2.60\pm 0.05)$ times the energy of
the first resonance which could indicate higher order processes of
two atoms tunnelling simultaneously.

\section{CONCLUSION}
We have studied the superfluid to Mott insulator transition in
one, two, and three spatial dimensions. The excitation spectra
were obtained by employing Bragg spectroscopy. We find the spectra
for the low-dimensional superfluid to differ significantly from
the three-dimensional case.

\end{document}